\documentclass[aps,pre,preprint,showpacs,amssymb]{revtex4}
\usepackage{amsmath,graphics}
\begin{document}
\title{The triangular Ising model with nearest-
 and next-nearest-neighbor couplings in a field}
\author{Xiaofeng Qian$^{1}$ and Henk W. J. Bl\"ote~$^{2,1}$} 
\affiliation{$^{1}$ Lorentz Institute, Leiden University,
  P.O. Box 9506, 2300 RA Leiden, The Netherlands}
\affiliation{$^{2}$Faculty of Applied Sciences, Delft University of
Technology, P.O. Box 5046, 2600 GA Delft, The Netherlands}
\date{\today}
\begin{abstract}
We study the Ising model on the triangular lattice
with nearest-neighbor couplings $K_{\rm nn}$,
next-nearest-neighbor couplings $K_{\rm nnn}>0$, and a magnetic field $H$.
This work is done by  means of finite-size scaling of numerical results
of transfer matrix calculations, and Monte Carlo simulations.
We determine the phase diagram and confirm the character of the critical
manifolds. The emphasis of this work is on 
the antiferromagnetic case $K_{\rm nn}<0$, but we also explore the 
ferromagnetic regime  $K_{\rm nn}\ge 0$ for $H=0$. For $K_{\rm nn}<0$
and $H=0$ we locate a critical phase presumably covering the whole range
$-\infty < K_{\rm nn}<0$. For $K_{\rm nn}<0$, $H\neq 0$  we locate a plane
of phase transitions containing a line of tricritical three-state Potts
transitions. In the limit $H \to \infty$ this line leads to a tricritical
model of hard hexagons with an attractive next-nearest-neighbor potential.
\end{abstract}
\pacs{05.50.+q, 64.60.Ak, 64.60.Cn, 64.60.Fr}
\maketitle
 
\section { Introduction
}\label{sec1}

\begin{figure}
\includegraphics{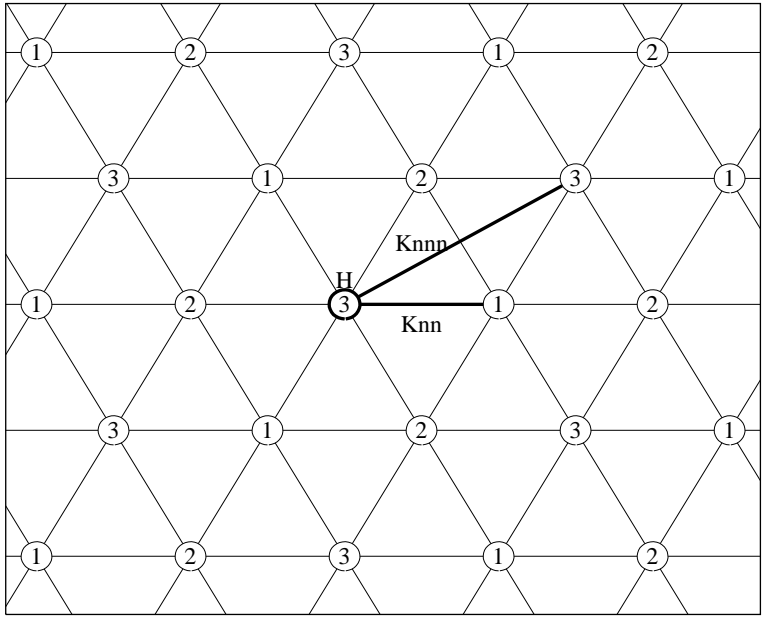}
\caption[xxx]
{The triangular lattice with nearest-neighbor couplings $K_{\rm nn}$, 
next-nearest-neighbor couplings $K_{\rm nnn}$ (examples of which are
shown as bold bonds), and a field $H$ (bold circle). 
The lattice is divided into three sublattices labeled $1$, $2$ and $3$. }
\label{fig1}
\end{figure}

The Ising model on the triangular lattice with nearest-neighbor couplings
$K_{\rm nn}$, next-nearest-neighbor couplings $K_{\rm nnn}$, and a
magnetic field $H$,  is defined by the reduced Hamiltonian 
\begin{equation}
{\mathcal H}/k_{\rm B}T = -  K_{\rm nn} \sum_{\langle{\rm nn}\rangle} s_i s_j 
- K_{\rm nnn} \sum_{[{\rm nnn}]} s_k s_l 
-H\sum_m s_m,  
\label{Ising}
\end{equation}
where $s_i=\pm 1$, and $\langle{\rm nn}\rangle$ and $[{\rm nnn}]$ indicate 
summations over all pairs of nearest neighbors and of next-nearest neighbors, 
respectively, as illustrated in Fig.~\ref{fig1}.   

This model, in particular the antiferromagnetic model 
($K_{\rm nn}< 0$), displays interesting behavior.
For $K_{\rm nnn}= 0,H=0$ the model has been solved exactly \cite{Hout}.
A ferromagnetic transition occurs at $K_{\rm nn}=\ln(3)/4$.
An antiferromagnetic ($K_{\rm nn}< 0$) mirror image of this transition
is {\rm absent}.
This is related to the fact that the triangular lattice is not bipartite.
However, at zero temperature, i.e., for $K_{\rm nn} \to - \infty$, 
the model displays a critical phase with algebraically decaying 
correlations \cite{ST}. This zero-temperature model can be exactly mapped on
a solid-on-solid (SOS) model \cite{BH}. Under renormalization, it is assumed 
to map on the 
Gaussian model \cite{N} and on the related Coulomb gas \cite{BNDL}.
The coupling constant $g_R$ of the Coulomb gas can thus be obtained
exactly as $g_R=2$ so that a number of critical exponents can be calculated. 
The  Ising temperature $T \propto -K_{\rm nn}^{-1}$
appears to be {\em relevant}: the critical state is destroyed for all
$T>0$. Commensurate-incommensurate transitions occur when finite
differences between the infinite nearest-neighbor couplings in the three
lattice directions are introduced \cite{BH,N}. 

Next we consider the case of $H=0$ and $K_{\rm nnn} \ne 0$. The mapping on 
the SOS model (and we may also assume this for the Coulomb gas) is still
valid for $K_{\rm nn} \to - \infty$ but, in the absence of an exact
solution, $g_R$ is no longer exactly known.
It has, however, been deduced \cite{N} that $g_R$ is an increasing
function of $K_{\rm nnn}$. The Coulomb gas analysis predicts that, for
sufficiently large $g_R$, the Ising temperature becomes irrelevant, 
so that the algebraic phase extends to nonzero temperatures. This analysis
also predicts that for even larger $g_R$ a phase transition to 
a flat SOS phase occurs, both at zero and at nonzero temperatures. 

Somewhat earlier, part of this scenario had already been described by
Landau \cite{DPL}. Via the lattice-gas representation of Eq.~(\ref{Ising}),
he used the connection with the XY model in the presence of  a six-state
clock-like perturbation, made earlier by Domany {\em et al.}~\cite{Dea}. 
He could
thus make use of their results \cite{Dea} for this model which allow for
the existence of a critical, XY-like phase in a nonzero range $K_{\rm nn}>0$.
Furthermore, Landau \cite{DPL} used the Monte Carlo method to verify the
existence and nonuniversal character of this critical phase for the case
of a fixed ratio $K_{\rm nnn}/K_{\rm nn}=-1$.

Another tool to study the model with nonzero next-nearest-neighbor couplings
$K_{\rm nnn}$ is provided by the transfer-matrix technique. A simplification
has been used in the latter approach: $K_{\rm nnn}$ was taken to be nonzero
only for four out of the six next-nearest neighbors \cite{KO,MKK,QD}. This
leads to a substantial simplification of the transfer matrix calculations,
but the resulting system lacks isotropy, so that applications of conformal
mappings become difficult. On this basis, limited evidence \cite{QD}
for the existence of the critical phase was reported; the limitation of
this evidence is obviously related to the lack of sixfold symmetry.

Next we consider the consequences of a nonzero field $H>0$.
On the basis of the relation with the Coulomb gas it has been
derived \cite{N} that, for $K_{\rm nn} \to - \infty$ and $K_{\rm nnn}=0$,
the magnetic field $H$ is {\em irrelevant}: the critical state is not
destroyed by a sufficiently small field $H\neq 0$. However, the magnetic
field tends to increase the Coulomb gas coupling constant $g_R$.
The field will become marginally relevant at $g_R=9/4$ and a
transition of the Kosterlitz-Thouless (KT) type or, in this context
more appropriate, of the roughening type is thus expected.
This transition separates the critical phase from a long-range
ordered phase, where the majority of the minus-spins have condensed on
one of the three sublattices of the triangular lattice. This prediction
has been confirmed \cite{BN,QWB} by means of numerical methods. 
The long-range ordered phase extends to nonzero temperature $T>0$ and
is separated from the disordered phase by a line of phase transitions
in the $(H,T)$ plane that belongs to the three-state Potts universality
class \cite{QWB,SA,KS,NK,TS}. 

Since the Ising model in a field can be mapped on a vertex model, and 
the critical manifolds of solvable vertex models are described by 
the zeroes of simple polynomials in the vertex weights \cite{FYWu}, it
may be assumed that also for the triangular lattice the critical line
in the $(H,T)$  is described by such a polynomial. This assumption was
recently refuted by Qian {\em et al.}~\cite{QWB}. The shape of the
critical line, as deduced from this assumption, was found to be 
inconsistent with the numerical evidence. They also found that the
renormalization ideas originally outlined by Nienhuis {\em et al.}~\cite{N}
could be applied to predict the shape of the critical line in the $(H,T)$
plane for small $T$. This shape was found to be consistent with their
numerical data for the critical line.

The aforementioned three-state Potts-like critical line is naturally part
of a critical surface extending to nonzero $K_{\rm nnn}$. The more involved
problem to find the phase diagram in the three-parameter ($H$, $K_{\rm nnn}$,
$K_{\rm nn}$) space has already been partly explored. On the basis of
renormalization arguments, Nienhuis {\em et al.}~\cite{N} obtained 
information about the shape of the critical surface in the
limit $H \to 0$. Landau \cite{DPL} performed Monte Carlo simulations for
a fixed ratio $K_{\rm nnn}/K_{\rm nn}=-1$. He determined the line of phase
transitions as a function of $H$ and noted that the three-state Potts
character along this line changes at a tricritical point beyond which     
the transition turns first-order.
  
In this work we verify the predictions in Ref.~\onlinecite{N} and
determine the critical values of $K_{\rm nnn}$ corresponding to 
several relevant values of the Coulomb gas coupling constant $g_R$, both    
for finite and infinite $K_{\rm nn}$. We verify the character of the
predicted critical phase at $H=0$.
We also study the critical phenomena associated with the introduction of 
a nonzero magnetic field and explore the full three-parameter phase diagram 
for $K_{\rm nnn} \geq 0$.

This paper is organized as follows. 
In Sec.~\ref{sec2}, we summarize our numerical methods which include
Monte Carlo algorithhms and the construction of a transfer matrix. We
define the observables that will be the subject of our numerical analysis. 
The study of the phase transitions of the triangular Ising model with
nearest- and next-nearest-neighbor couplings in a zero field is presented
in Sec.~\ref{sec3}, and in Sec.~\ref{sec4} we describe our results
for a nonzero magnetic field; we conclude with a discussion in
Sec.~\ref{sec5}. 

\section{Numerical methods} \label{sec2}
\subsection {Transfer-matrix calculations} \label{TM}
Most of the the transfer-matrix calculations were performed for $T>0$ so 
that we had to use a binary representation for the Ising spins, leading to 
a transfer matrix of size $2^L \times 2^L$ for a system with finite size $L$.
For $T=0$ one can use a simplified transfer matrix of a smaller size \cite{BN}. 
We define the spin lattice  on the surface of a cylinder, and take the 
transfer direction perpendicular to a set of  nearest-neighbor edges. 
The lattice is divided into three sublattices denoted as $1$, $2$ and $3$, 
respectively, as shown in Fig.~\ref{fig1}. Nearest-neighbor interactions 
occur only between different sublattices and next-nearest-neighbor interactions 
occur within the same sublattice. 

To enable  calculations for system as large as possible, a sparse matrix 
decomposition has been used. This leads to a very significant reduction
of the required computer time and memory. The transfer matrices 
are defined in \cite{BN,QWB} for the nearest-neighbor model. 
Here we modify the transfer matrix to include all next-nearest-neighbor
interactions. This makes it necessary to code two (instead of one) layers
of spins as the transfer matrix index.  Finite-size calculations with
$L$ multiples of 6 up to $L=24$ were performed.
The maximum finite size $L=24$ corresponds to a cylinder with a 
circumference of only $12$ nearest-neighbor bonds.   

The magnetic correlation function along the coordinate $r$ in the
length direction of the cylinder is defined as:
\begin{equation}
g_m(r) = \langle s_{\rm 0}s_r\rangle.
\end{equation}
At large $r$, this correlation function decays exponentially with a
characteristic length scale $\xi_m$ that depends on $K_{\rm nn}$,
 $K_{\rm nnn}$, $H$, and $L$:
\begin{equation}
g_m(r) \propto {\rm e}^{-r/\xi_m (K_{\rm nn},K_{\rm nnn},H,L)}
\end{equation}
which can be calculated from the largest two eigenvalues $\lambda_0$ 
and $\lambda_1$ of the transfer matrix:
\begin{equation}
\xi_m^{-1}(K_{\rm nn},K_{\rm nnn},H,L) =
\frac{1}{2 \sqrt{3}} \ln(\lambda_0/\lambda_1),
\end{equation}
where the factor $2\sqrt{3}$ is a geometric factor for two layers of spins. 
For the calculation of $\xi_m$, we make use of the symmetry of the eigenvectors 
associated with $\lambda_0$ and $\lambda_1$. The leading eigenvector 
(for $\lambda_0$) is invariant under a spatial inversion. 
In contrast, the second eigenvector is antisymmetric under inversion.

The theory of conformal invariance \cite{Cardy} relates $\xi_m$ on 
the cylinder with the 
magnetic scaling dimension $X_m$ (one half of the magnetic correlation function 
exponent $\eta$). This exponent may be estimated as 
\begin{equation}
X_m(K_{\rm nn},K_{\rm nnn},H,L) = \frac{L}{2 \pi \xi_m(K_{\rm nn},K_{\rm nnn},H,L)}.
\label{XML}
\end{equation}
 Asymptotically for a critical model with large $L$ we have
\begin{equation}
X_m(K_{\rm nn},K_{\rm nnn},H,L) \simeq X_m,
\label{XM}
\end{equation}
where $X_m=1/(2g_R)$ in the language of Coulomb gas.
This equation allows us to estimate $X_m$ numerically and thus to obtain 
evidence about the universality class of the model. Or, if the universality 
class, and thus $X_m$, are considered known, Eq.~(\ref{XM}) and be used to 
determine the critical surface, e.g. to solve for  $K_{\rm nnn}$ 
for given values of $K_{\rm nn}$, $H$ and $L$. 
As a consequence of corrections to scaling, the solution will not
precisely coincide with the critical point. The effects of
an irrelevant scaling field $u$ and a small deviation $t$ with
respect to the critical value of $K_{\rm nn}$, or $K_{\rm nnn}$, or $H$
 are expressed by
\begin{equation}
X_{m}(K_{\rm nn},K_{\rm nnn},H,L) =  X_m + au L^{y_i} + b t L^{y_t} + \cdots,
\label{xhs}
\end{equation}
where $a$ and $b$ are unknown constants, $y_i$ is irrelevant exponent
and $y_t$ is temperature exponent. 
For the solution of the equation $X_{m}(K_{\rm nn},K_{\rm nnn},H,L) =  X_m$
we thus have $au L^{y_i} + b t L^{y_t} \approx 0$, so that we expect
corrections proportional to $L^{y_i-y_t}$ in the critical point estimates.
For instance, for three-state Potts universality one has $y_t=6/5$ and
$y_i=-4/5$ so that the leading finite-size dependence of the estimated
critical points is as $L^{-2}$. This knowledge is helpful for the
extrapolation to the actual $L=\infty$ critical point.

In addition to $\xi_m$, it is possible to determine a second correlation 
length $\xi_t$ describing the exponential decay of the energy-energy 
correlation function. It is associated with a third eigenvalue $\lambda_2$ 
of the transfer matrix with an eigenvector that is symmetric under a spatial
inversion, just as the one with eigenvalue $\lambda_0$. The pertinent
eigenvalue is thus solved by means of orthogonalization with respect to
the first eigenvector.  In analogy with the case of the magnetic 
correlation length we can use the third eigenvalue $\lambda_2$ to 
estimate the temperature-like scaling dimension $X_t$ as 
\begin{equation}
X_t(K_{\rm nn},K_{\rm nnn},H,L) = \frac{L}
{2 \pi \xi_t(K_{\rm nn},K_{\rm nnn},H,L)},
\label{XTL}
\end{equation}
where $\xi_t=\frac{1}{2\sqrt3}\ln(\lambda_0/\lambda_2)$.
At criticality, it behaves for large $L$ as:
\begin{equation}
X_t(K_{\rm nn},K_{\rm nnn},H,L) \simeq X_t.
\label{XT}
\end{equation}
Combining Eqs.~(\ref{XM}) and (\ref{XT}), we can solve for two
unknowns simultaneously, using the known \cite{BNDL}
values of the tricritical three-state Potts model, namely $X_m=2/21$
and $ X_t=2/7$. In this way, we can estimate the tricritical point 
($K_{\rm nnn},K_{\rm nn}$) for a given $H$. The corrections
can be argued to be proportional to $L^{y_i-y_{t2}}$ where $y_{t2}=4/7$ 
and $y_i=-10/7$, i.e., the corrections decay as $L^{-2}$.

\subsection{Monte Carlo simulations}
Since transfer-matrix calculations are, although highly accurate,
restricted to small systems, we have also written Monte Carlo algorithms
for the present model. To obtain good statistical accuracies we included
not only a Metropolis algorithm, but also a Wolff and a geometric cluster
algorithm.  Which algorithm is used 
depends on the location in the phase diagram. The Wolff algorithm  
is applicable in only the case of zero magnetic field. The geometric 
algorithm \cite{HB} conserves the magnetization and was therefore used 
in combination with the Metropolis algorithm. This combination was found 
to work faster than the Metropolis method, but the gain in efficiency
depends on the position in the  three-parameter space.

Several quantities were sampled using  these algorithms in order to
explore the phase diagram.
First we define the uniform magnetization as $m\equiv L^{-2} \sum_k s_k$ 
which tends to $\pm 1/3$ in the long-range ordered antiferromagnetic or
`flat' phases, and to zero in the disordered (paramagnetic) phase.
From its moments we define the magnetic Binder ratio as
\begin{equation}
Q_m  = \frac {\langle m^2\rangle^2}{\langle m^4\rangle}.
\end{equation}
Next, we consider the three-state Potts-like order parameter or, in the 
language of the present Ising model, the three sublattice magnetizations.
We denote the magnetization density of sublattice $i$ ($i=1$, 2, or 3)
as $m_i$. On the basis of the staggered magnetizations we write the
variance of the Potts order parameter as
\begin{equation}
m_{\rm s}^2 = m_1^2 +  m_2^2  +  m_3^2 -  m_1m_2 -  m_2m_3 -  m_3m_1 
\end{equation}
and the corresponding dimensionless ratio as
\begin{equation}
Q_{\rm s}  = \frac {\langle m_{\rm s}^2\rangle ^2}{\langle m_{\rm s}^4\rangle}.
\end{equation}
At criticality, the quantities $Q_m$ and $Q_{\rm s}$ scale as a constant plus
irrelevant corrections, i.e., they converge to a constant as $L$ increases.
This property can be used for the determination of critical points.

\section{Numerical results for zero field}\label{sec3}
We restrict this work to ferromagnetic next-nearest-neighbor interactions 
($K_{\rm nnn}>0$). First, we consider the Ising model in a zero field 
($H=0$), and study the phase diagram in ($K_{\rm nnn},K_{\rm nn}$) plane. 
We distinguish the cases $K_{\rm nn}>0$ and $K_{\rm nn}<0$.  

\subsection{ Results for the ferromagnetic transition ($K_{\rm nn}>0$)}
\label{sec3sub1}
For the Ising model we have $X_m=1/8$ so that at criticality 
we expect that asymptotically for large $L$
\begin{equation}
X_ m(K_{\rm nn},K_{\rm nnn},0,L)\simeq \frac{1}{8}
\end{equation}
from which one can estimate critical points e.g. by solving for 
$K_{\rm nnn}$ at a given value of  $K_{\rm nn}$ or vice versa. 
In certain cases,  critical points can be  determined 
accurately by extrapolating to $L=\infty$ . 
For instance, for $K_{\rm nnn}=0$ we obtain the  critical value of 
the nearest-neighbor coupling $K_{\rm nn}=0.2746528(10)$, 
which is consistent with the exact result $K_{\rm nn}= \ln(3)/4$. 
The results are shown in Fig.~\ref{fig2}.

We also checked that,
at the decoupling point ($K_{\rm nn}= 0$)
the critical value of 
the next-nearest-neighbor coupling $K_{\rm nnn}$ 
equals the exact value $\ln(3)/4$ . 
The three  sublattices, which are also triangular lattices,
become independent  at the decoupling point. 

\subsection { Results for the antiferromagnetic region ($K_{\rm nn} < 0$)}
At finite $K_{\rm nn}<0$ and small $K_{\rm nnn}>0$, the model is obviously
disordered. As described in the Introduction, with
increasing $K_{\rm nnn}$ the model is expected to undergo:
1) a Kosterlitz-Thouless transition to a critical phase 
at the point where the Coulomb gas coupling reads $g_R=4$, and
the corresponding value of the magnetic dimension is 
$X_m=1/(2g_R)=1/8$;
2) a roughening transition to a flat phase,
and the corresponding value of the magnetic dimension is thus 
$X_m=1/18$ at $g_R=9$.
We have solved $K_{\rm nnn}$ from Eq.~(\ref{XM}) for these two values of
$X_m$, at several fixed 
values of $K_{\rm nn}$. The results were extrapolated to $L=\infty$
by means of three-point fits involving a constant (the estimated value
of $K_{\rm nnn}$) plus a finite-size correction involving a free 
exponent. The final estimates are included in the phase diagram 
Fig.~\ref{fig2}. They suggest that the two boundaries of the 
critical phase merge at the decoupling point $K_{\rm nn}=0$.
Our numerical results include a few special points at zero temperature
($K_{\rm nn} \to - \infty$). In the renormalization scenario, their 
meaning is as follows:
\begin{enumerate}
\item
For $g_R=9/4$ we obtain $K_{\rm nnn}=0.0185$ (4). This is where the
line of roughening transitions in the ($K_{\rm nnn},H$) plane meets the
$K_{\rm nnn}$ axis.
\item
For $g_R=3$ we obtain $K_{\rm nnn}=0.0667$ (2). This is where the line
of three-state Potts transitions in the plane perpendicular to the
$K_{\rm nnn}$ axis comes in as a straight line with a
nonzero, finite slope as argued in Ref.~\onlinecite{QWB}.
\item
For $g_R=4$ we obtain $K_{\rm nnn}=0.1179$ (2). This is where the KT-like
line in the $(K_{\rm nnn},K_{\rm nn})$ plane meets the $K_{\rm nnn}$ axis.
\item
For $g_R=9$ we obtain $K_{\rm nnn}=0.226$ (2). This is where the line of
roughening transitions in the $(K_{\rm nnn},K_{\rm nn})$ plane meets the
$K_{\rm nnn}$ axis. This point corresponds with an actual phase transition
on the $K_{\rm nnn}$ axis. We note that, in cases 1 and 3, the $K_{\rm nnn}$
axis meets with other lines of phase transitions. However, phase transitions 
do not occur at points 1 and 3 because the critical amplitudes vanish on 
the $K_{\rm nnn}$ axis.
\end{enumerate}
\subsection{Shape of the critical lines for small $|K_{\rm nn}|$}
On the basis of an argument due to van Leeuwen \cite{vL}, the scaling 
behavior of $K_{\rm nn}$ near the decoupling point 
($K_{\rm nnn}=\ln(3)/4$, $K_{\rm nn}=0$), is governed by a new critical 
exponent $y_a=7/4$. This exponent thus determines the shape of the critical 
lines for small  $|K_{\rm nn}|$ according to
\begin{equation}
K_{\rm nn}  \propto \left(\frac{\ln 3}{4} - K_{\rm nnn}\right)^{7/4}.
\label{shape}
\end{equation}

One can find the critical exponent $y_a$ exactly from the known properties
of the magnetic correlation function of the critical Ising model. The
spin-spin correlation behaves as
\begin{displaymath}
g_m(r) \propto r^{-2X_m},
\end{displaymath}
where $X_m=1/8$ for the 2D Ising model. 
This also applies to the decoupling point where the model decomposes 
in three independent sublattices. This determines the scaling behavior of a 
four-spin correlation function involving spins in different sublattices 
in the limit of $K_{\rm nnn} \to 0$
\begin{equation}
g_a(r) = \langle s_{00}s_{01}s_{r0}s_{r1}\rangle
=[g_m(r)]^2\propto r^{-4X_m},
\label{subs}
\end{equation}
where $s_{00}$ and $s_{01}$ are nearest-neighbor spins belonging to 
different sublattices, say sublattices 1 and 2. The same applies to 
the pair ($s_{r0}$, $s_{r1}$) at a distance $r$. 
Eq.~(\ref{subs}) describes the energy-energy correlation associated with
$K_{\rm nn}$.  Its power-law decay is thus expressed by 
\begin{equation}
g_a(r) \propto r^{-2X_a},
\label{scal}
\end{equation}
where $X_a$ is the scaling dimension of the nearest-neighbor energy density.
Comparing the two Eq.~(\ref{subs}) and Eq.~(\ref{scal}), we conclude that
$X_a =  2X_m =1/4$ and  $y_a=7/4$.

We verify Eq.~(\ref{shape}) by plotting $K_{\rm nn}$ versus
$[\ln(3)/4 - K_{\rm nnn}]^{7/4}$ for the ferromagnetic critical
line in Fig.~\ref{figfm}, and for the two lines containing the algebraic 
phase in the antiferromagnetic region in Fig.~\ref{figfaf}. 
In all these cases we find  approximate 
linear behavior near the decoupling point which confirms 
the predicted value of $y_a$.

\subsection{The algebraic phase}
The renormalization scenario predicts that, in the algebraic phase 
the estimates of  $X_m$, as obtained from Eq.~(\ref{XML}), will converge
to a $K_{\rm nnn}$-dependent limit when the finite size $L$ increases. 
However, in the disordered and flat phases, the system will renormalize
away from the nonuniversal fixed line, and the data for $X_m$ are
therefore predicted to fan out for different values of $L$.
We calculated  $X_m$ by solving Eq.~(\ref{XM}) 
in a suitable range of $K_{\rm nnn}$ at fixed values of $K_{\rm nn}$, 
namely $K_{\rm nn}=-\infty$, $-0.6$, $-0.4$, $-0.2$ and $-0.1$. 
These results confirm the renormalization predictions, as illustrated
in Figs.~\ref{figT0} and \ref{figk06}. 
Fig.~\ref{figT0} shows that, for $K_{\rm nn}=-\infty$ and $H=0$, 
the data of $X_m$ converge to a $K_{\rm nnn}$-dependent constant 
in a range of $K_{\rm nnn}$ from zero to $K_{\rm nnn}=0.226$ (2)
as determined above. This confirms that 
for $H=0$, $K_{\rm nn}=-\infty$ the system indeed remains critical 
until $K_{\rm nnn}$ induces a transition to a flat phase.
In contrast, Fig.~\ref{figk06} indicates that for nonzero temperature 
the critical phase starts at a positive value of $K_{\rm nnn}$.
Fig.~\ref{figk06i} shows the inverse of $X_m$ and provides a 
clearer picture of the transition at the large $K_{\rm nnn}$ side. 
We have numerically calculated the average slopes $S_L$ of the 
finite-size curves in intervals specified in Table I, and fit them 
as follows: 
\begin{equation}
S_L=S_{\infty}+ a L^{y_c}+\cdots,
\label{slope}
\end{equation} 
where $S_{\infty}$ is constant, and $y_c$ denotes the exponent of the
leading finite-size correction.
Results listed in Table~\ref{table1} indicate that the finite-size
dependence of the slopes is governed by a {\em negative} 
exponent $y_c$ of $L$, which indicates that the slope $S_L$ converges to
a constant for $L \to \infty$, as expected in the critical range.  

In order to provide independent confirmation of the algebraic phase, we 
also used the Monte Carlo method.  Simulations were done for $L\times L$
systems of size $L=24$, 36, 48, and 60.  Examples of the results for
$Q_{\rm s}$ and $Q_m$ are given in Fig.~\ref{figqs} and Fig.~\ref{figqm}
respectively, as a function of $K_{\rm nnn}$, for $K_{\rm nn} = -0.2$.
These data behave similarly as those for $X_m$, and show good apparent
convergence to a nonuniversal, $K_{\rm nnn}$-dependent constant in the
pertinent range.
Note that the curves for $Q_{\rm s}$ display intersections near
$K_{\rm nnn} \approx 0.207$, and those for $Q_m$ near
$K_{\rm nnn} \approx 0.245$, apparently at different sides of the
algebraic phase as shown in Fig.~\ref{fig2}. We interpret these
intersections, i.e., solutions of Eq.~(\ref{XM}) coinciding for different
$L$, as the cancellation of the leading two $L$-dependent terms. Such 
terms are likely associated with 1) the corrections as naturally
associated with irrelevant fields in the algebraic phase;
and 2) the `fanning-out' phenomenon
mentioned above. It appears that the first types of corrections in
$Q_{\rm s}$ and $Q_m$  are of a different sign. 

\section{Results for nonzero field} \label{sec4}
In view of the Ising character of (\ref{Ising}), we restrict ourselves 
to $H\ge 0$ without loss of generality.
The phase diagram without next-nearest-neighbor interactions, i.e.,
in the ($H,K_{\rm nn}$) plane has already been determined by Qian
{\em et al.}~\cite{QWB}, with special emphasis on the limit
$K_{\rm nn} \to -\infty$. In that limit, a roughening-type transition
is located \cite{BN,QWB} near $H=0.266$. As mentioned above, the
algebraic phase becomes less stable against perturbation by $H$ when
$K_{\rm nnn}$ increases, and the algebraic phase in the ($K_{\rm nnn},H$)
plane shrinks to zero at $g_R=9/4$ which corresponds, as mentioned above,
to $K_{\rm nnn}=0.0185$.

The line connecting the two points $(K_{\rm nnn},H)$ =(0,0.266) and
(0.0185,0) is a line of roughening transitions separating the algebraic
and the ordered phases. The renormalization description implies that this
line is a straight line when expressed in the scaling fields. In view of the
proximity of both numerically determined points, we expect an almost
straight line in the $(K_{\rm nnn},H)$ plane. 
The connection of the three-state Potts transition line and the roughening
transition point in ($H,K_{\rm nn}$) plane has been analytically investigated
by Qian {\em et al.} using renormalization arguments. Their analysis indicates
that the roughening transition at  $H=0.266$ is the end point of the Potts
transition line in ($H,K_{\rm nn}$) plane for $T \downarrow 0$. 
Their result applies similarly to other points on the line of roughening 
transitions. We thus believe that this whole line serves as a frontier of
the Potts critical surface, as well as the part of the $K_{\rm nnn}$ axis
with $g_R$ between $9/4$ and 4 as determined in Sec.~\ref{sec3sub1}.

Since three-state Potts universality implies $X_m=2/15$ at criticality,
we expect that asymptotically for large $L$
\begin{equation}
X_ m(K_{\rm nn},K_{\rm nnn},H,L)\simeq \frac{2}{15}
\end{equation}
from which one can estimate critical points by solving for one of the
three variables $(K_{\rm nn},K_{\rm nnn},H)$ for specified values
of the other two, and subsequent extrapolation to $L=\infty$.
We thus calculated critical points on several lines at fixed values of
$H$. The results are shown as lines connecting these points in
Fig.~\ref{fig7}. In order to zoom in on the connection of the
three-state Potts transition surface and the transition lines in the
($K_{\rm nnn},K_{\rm nn}$) plane, we have also estimated critical
values of $H$ at fixed values of $K_{\rm nn}$, for a suitably chosen
range of $K_{\rm nnn}$.  Results for $K_{\rm nn}=-0.8,-0.1,-0.15$
are included in Fig.~\ref{fig7}. They fit well with the qualitative
predictions for the shape of the critical surface \cite{N} for small $H$.
Furthermore, our data for the critical points at $K_{\rm nnn}=0.0667$,
corresponding with $g_r=3$, agree with the linear behavior as mentioned
in Sec.~\ref{sec3sub1}.

Our results confirm that, when the next-nearest-neighbor coupling
$K_{\rm nnn}$ becomes sufficiently strong, the transition from the
disordered phase to the ordered phase changes character at a
tricritical line, beyond which the transition turns first-order. 
We have located the tricritical line using transfer-matrix calculations.
By solving Eqs.~(\ref{XM}) and (\ref{XT}) simultaneously for $K_{\rm nn}$
and $K_{\rm nnn}$ at specified values of $H$, we obtain results shown
in Table \ref{table2}, and included in  Fig.~\ref{fig7}.
In comparison with transfer-matrix calculations involving only $X_m$,
the memory requirements are somewhat larger. As a consequence only 3
values of $L$ up to 18 could be used. But we found that finite-size 
corrections are relatively small, and we are confident
that the tricritical line is well determined. 

For sufficiently large fields $H$, triangles may contain at most
one minus-spin and the tricritical line approaches 
a tricritical lattice-gas limit. In this limit the nearest-neighbor
coupling and the field satisfy a linear relation 
\begin{equation}
K_{\rm nn}=- \frac{H}{6}+ C.
\label{LG}
\end{equation}
As illustrated in Fig.~\ref{trifig}, the numerical data fit this
expression well, except at small $H$.
In order to obtain a satisfactory fit to the numerical data
for $H\geq 1$, we added terms proportional to $e^{-2H/3}$ and 
$e^{-4H/3}$ to Eq.~(\ref{LG}). This fit 
yielded $C=-0.01481$ (5). A similar fit without a term
proportional to $H$ yielded $K_{\rm nnn}=0.23514$ (7) for the 
tricritical lattice gas limit.

We have used Monte Carlo simulations to determine the location of the
sheet of first-order transitions at $K_{\rm nnn}=0.3$. We found that,
depending on $K_{\rm nn}$ and $H$, a randomly initialized system 
evolved to a phase either largely magnetized, or resembling one of the 
three ordered Potts states. The threshold values between these two
regimes are shown by the heavy dashed lines in Fig.~\ref{fig7}.
They fit smoothly with the results obtained in the critical range and
for the tricritical line.

\section{Discussion}\label{sec5} 

We have determined the phase diagram of the model Eq.~(\ref{Ising}) 
for $K_{\rm nnn} \ge 0$.  We locate a surface of phase transitions.
This surface divides into  a three-state Potts-like critical sheet
and a first-order part. The two parts are separated by a tricritical
line.
While the determination of tricritical line becomes less accurate for
small $|K_{\rm nn}|$, our data suggest that it spans the whole range 
$-\infty<K_{\rm nn}<0$. This is in agreement with the minimal
renormalization scenario in which the tricritical line is a flow line 
leading directly from the decoupling point to the tricritical fixed
point.

For $H \to \infty$, minus-spins are excluded on nearest-neighbor
sites and the the substitution $\sigma_i=(1-s_i)/2$ reduces the model
to a hard-hexagon lattice gas described by
the reduced Hamiltonian
\begin{equation}
{\mathcal H}_{hh}/k_{\rm B}T =
  V_{\rm nn}  \sum_{\langle{\rm nn}\rangle} \sigma_i  \sigma_j 
+ V_{\rm nnn} \sum_{[{\rm nnn}]} \sigma_k  \sigma_l
- \mu \sum_m \sigma_m \; ,
\label{hhex}
\end{equation}
where the site variables assume values $\sigma_i=0,1$ 
and $V_{\rm nn} \to \infty$ so that nearest-neighbor exclusion
applies. The chemical potential of the lattice-gas particles depends
on the Ising parameters as $\mu= -12K_{\rm nn}-12K_{\rm nnn}-2H$, and 
the next-nearest-neighbor potential as $V_{\rm nnn}=-4K_{\rm nnn}$.
For $V_{\rm nnn}=0$ this model reduces to Baxter's hard-hexagon 
lattice gas \cite{Bax}.
According to the analysis presented in Sec.~\ref{sec4}, the
tricritical line persists in the lattice-gas limit. The Ising
parameters $C$ and $K_{\rm nnn}$ determine the tricritical parameters
of the lattice-gas as $\mu=-2.644$ (1) and $V_{\rm nnn}=-0.9406$ (3).
Our findings may be compared with those of Verberkmoes and Nienhuis
\cite{VN} for a model with $V_{\rm nnn}= 0$ but including additional
smaller hexagons.
They also report a tricritical point, attributed to an effective
attraction between the hard hexagons, induced by entropic effects
associated with the small hexagons.

An Ising-like tricritical point is known to occur also in the 
analogous case of the hard-square lattice gas \cite{Bax,Bb,H}. Our result
thus confirms that tricriticality is a generic property of hard-core
lattice gases with attractive next-nearest-neighbor interactions.

Since we do not doubt the universality class of the tricritical line, 
we have not explicitly determined its critical exponents. However, we
remark that the fast apparent convergence of the estimated
tricritical points
confirms that the values of the Potts tricritical exponents $X_m$ and
$X_t$, as used to solve Eqs.~(\ref{XM}) and (\ref{XT}), do indeed apply.

\begin{figure}
\includegraphics{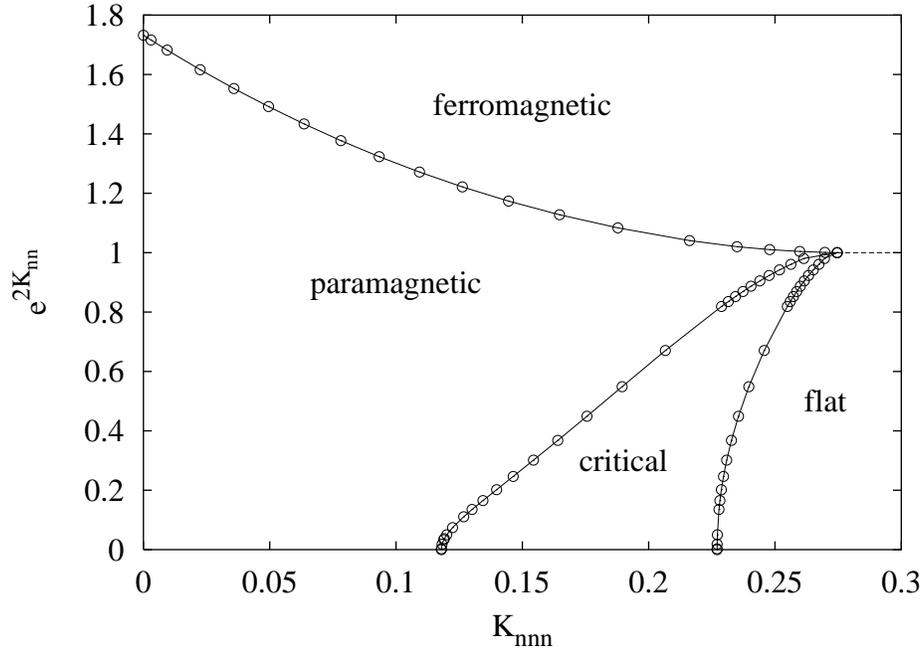}
\caption[xxx]
{Three lines of phase transitions in the $(K_{\rm nnn},K_{\rm nn}$ plane.
The numerically determined data points are shown as circles.
The upper line displays the ferromagnetic critical line for $K_{\rm nn}>0$.
For $K_{\rm nn}<0$ there are two more lines which represent the boundaries
of a  critical phase which resembles the low-temperature phase of the XY
model. The two lines appear to meet at a single point, the decoupling point,
at $K_{\rm nn}=0$. The right hand critical line marks a roughening
transition to a flat SOS phase, the left hand line a KT-like transition
between the disordered and the critical phases. The numerical errors in
the ferromagnetic region are much smaller than the size of the symbols;
for the remaining data they are difficult to estimate but believed to be
at most of the same order as the symbol size. }
\label{fig2}
\end{figure}

\begin{figure}
\includegraphics{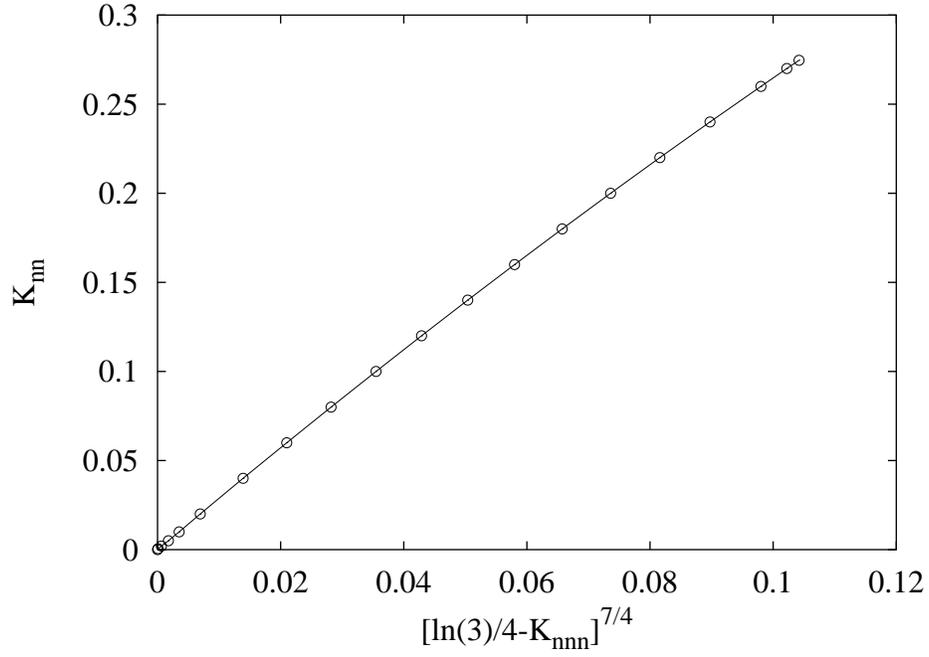}
\caption[xxx] 
{The ferromagnetic critical line, plotted as $K_{\rm nn}$ versus 
$[\ln(3)/4-K_{\rm nnn}]^{7/4}$. The approximate linear behavior confirms 
that the exponent $y_a$ associated with $K_{\rm nn}$ obeys the theoretical
prediction $y_a=7/4$. The estimated errors are smaller than the symbol size.}
\label{figfm}
\end{figure}

\begin{figure}
\includegraphics{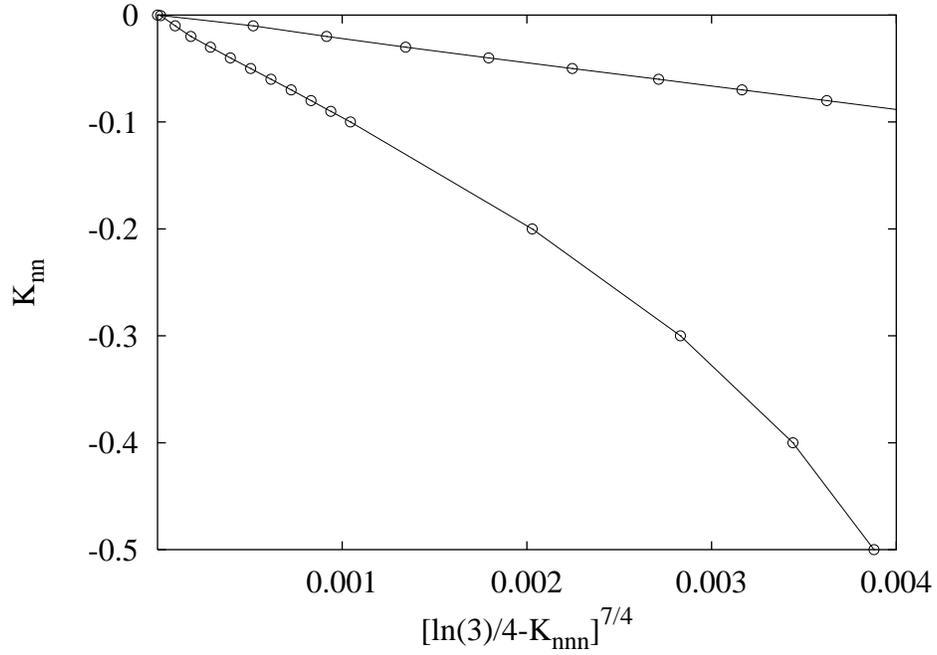}
\caption[xxx]
{Antiferromagnetic ($K_{\rm nn}<0$) critical lines near the decoupling point.
The numerical results (circles) are plotted as $K_{\rm nn}$ versus
$[\ln(3)/4 - K_{\rm nnn}]^{7/4}$. The approximate linear behavior at
small $|K_{\rm nn}|$ confirms that the exponent associated with the
scaling of $K_{\rm nn}$ obeys the theoretical prediction $y_a=7/4$.
The estimated errors in the data points are at most of the same order
as the symbol size.}
\label{figfaf}
\end{figure}

\begin{figure}
\includegraphics{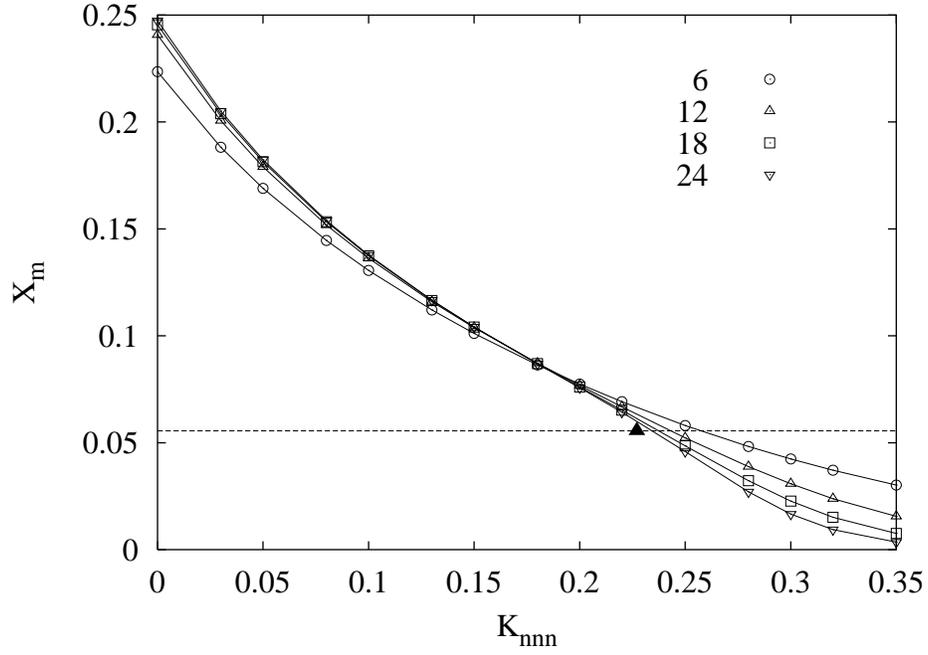}
\caption[xxx]
{Finite-size estimates of the magnetic scaling dimension $X_m$ versus
next-nearest-neighbor coupling $K_{\rm nnn}$ at $K_{\rm nn}=-\infty$. For
clarity we include four lines connecting data points for system sizes $L=6$,
12, 18, 24 respectively. The dashed line indicates the special value 
$X_m=1/18$, and the black triangle shows the estimated critical 
value of $K_{\rm nnn}$ for $K_{\rm nn}\to -\infty$.}
\label{figT0}
\end{figure}

\begin{figure}
\includegraphics{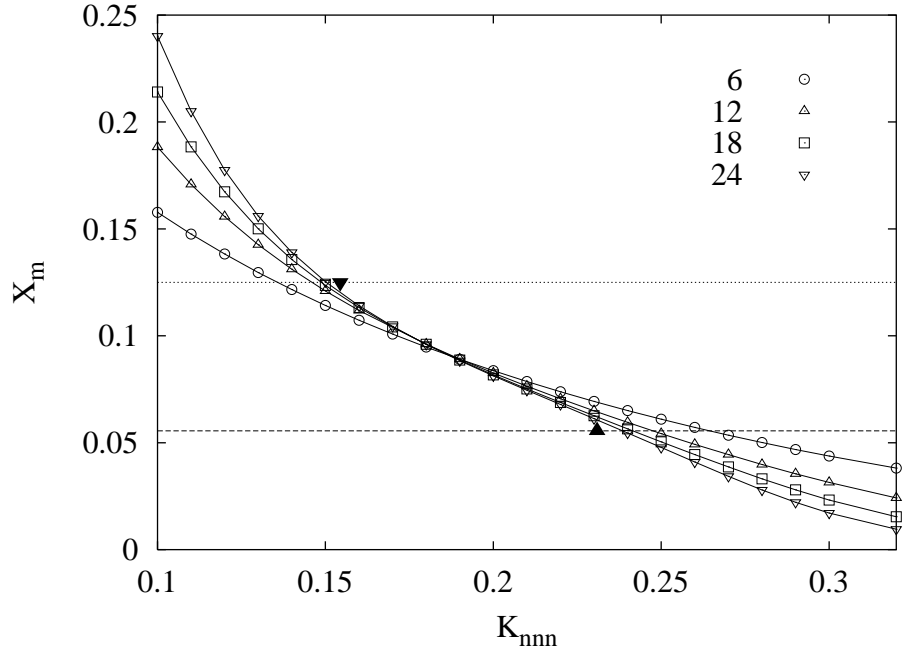}
\caption[xxx]
{Finite-size estimates of $X_m$ versus $K_{\rm nnn}$ at $K_{\rm nn}=-0.6$.
For clarity we include four lines connecting data points for system sizes
$L=6$, 12, 18, 24 respectively.  The dotted and dashed lines indicate the
special values $X_m=1/8$ and $X_m=1/18$ respectively.
The two black triangles show the estimated critical values of 
$K_{\rm nnn}$ at $K_{\rm nn}=-0.6$.}
\label{figk06}
\end{figure}

\begin{figure}
\includegraphics{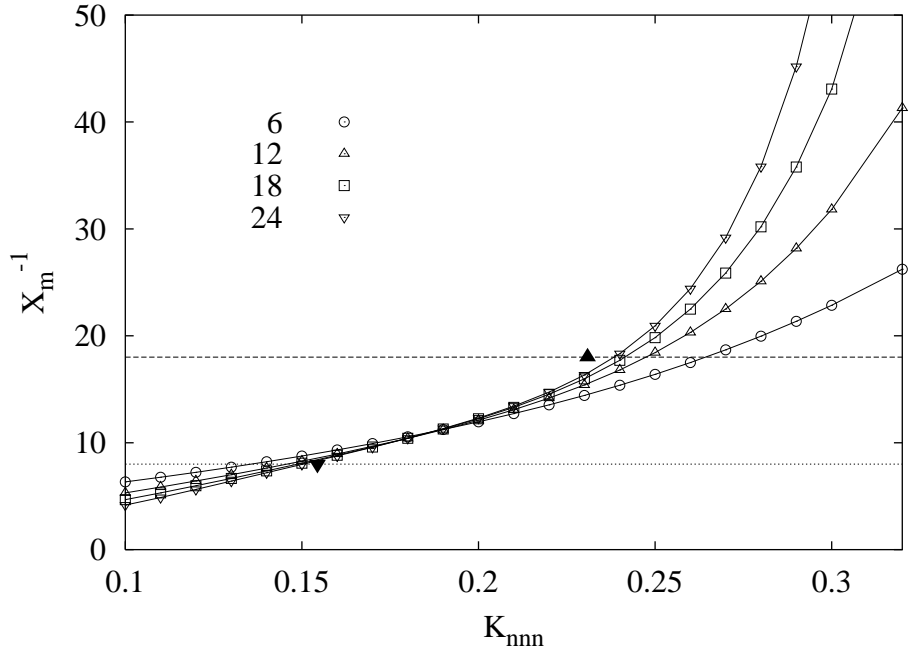}
\caption[xxx]
{Finite-size estimates of the inverse magnetic scaling dimension $X_m^{-1}$
versus next-nearest-neighbor coupling $K_{\rm nnn}$ at $K_{\rm nn}=-0.6$. 
The meaning of the lines and symbols are the same as in Fig.~\ref{figk06}.
The phase transition to flat phase is clearly visible in this figure.}
\label{figk06i}
\end{figure}

\begin{figure}
\includegraphics{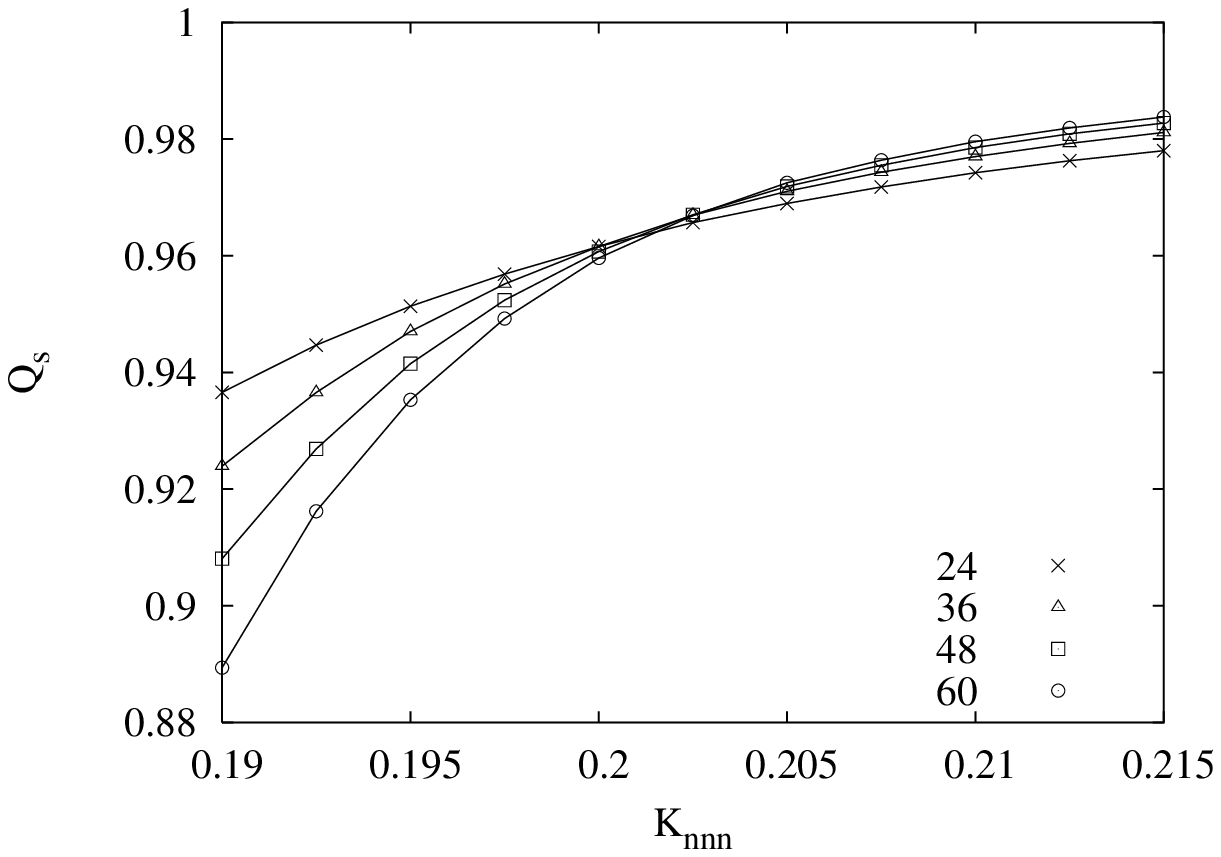}
\caption[xxx]
{Dimensionless amplitude ratio $Q_{\rm s}$ versus $K_{\rm nnn}$ at
$K_{\rm nn}=-0.2$. Intersections are found to occur near the transition 
point between the disordered and the algebraic phases. The four lines 
connecting the data points represent, with increasing slope, system 
sizes $L=24$, 36, 48, and 60, respectively. The numerical uncertainty
margins are much smaller than the size of the data points.}
\label{figqs}
\end{figure}

\begin{figure}
\includegraphics{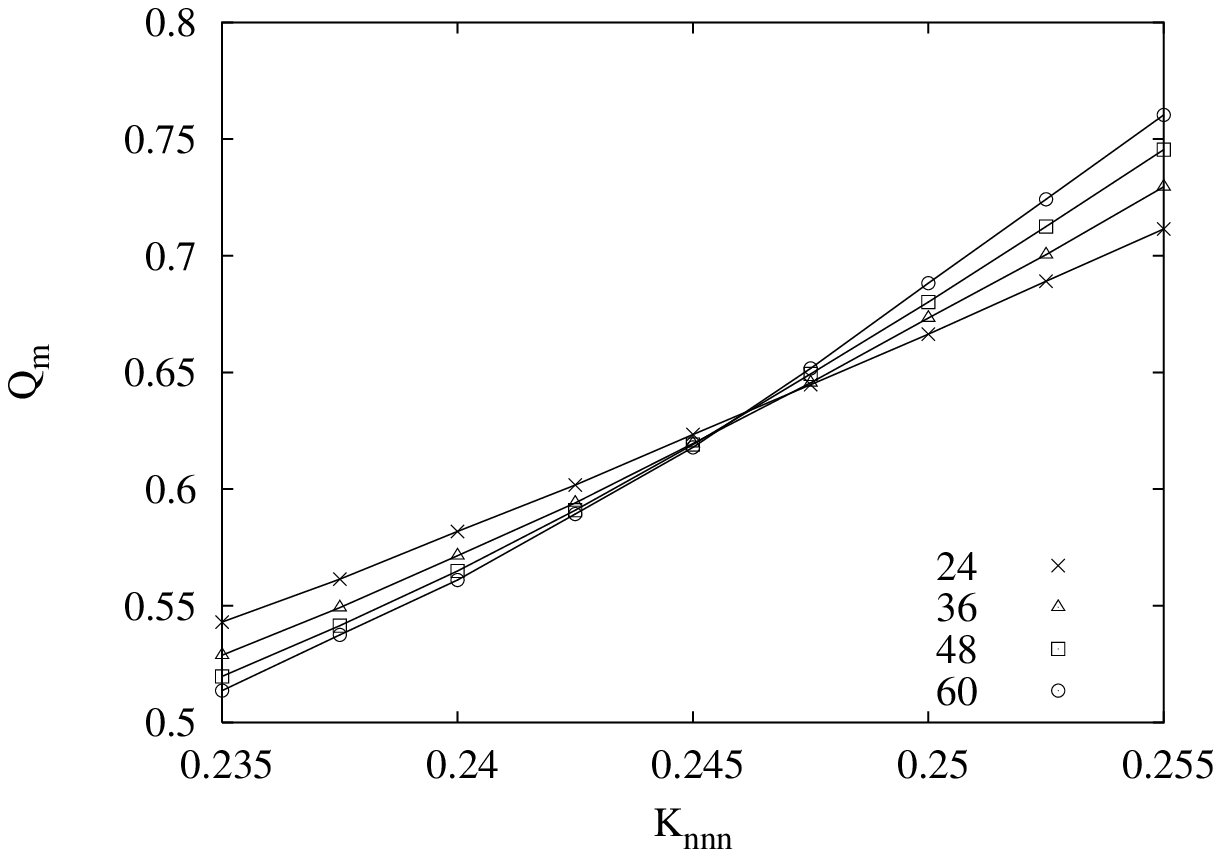}
\caption[xxx]  
{Dimensionless amplitude ratio $Q_m$ versus $K_{\rm nnn}$ at 
$K_{\rm nn}=-0.2$. Intersections are found to occur near the transition
point between the algebraic and the flat SOS  phases. The four lines
connecting the data points represent, with increasing slope, system
sizes $L=24$, 36, 48, and 60, respectively. The numerical uncertainty
margins are much smaller than the size of the data points.}
\label{figqm}
\end{figure}

\begin{figure}
\includegraphics{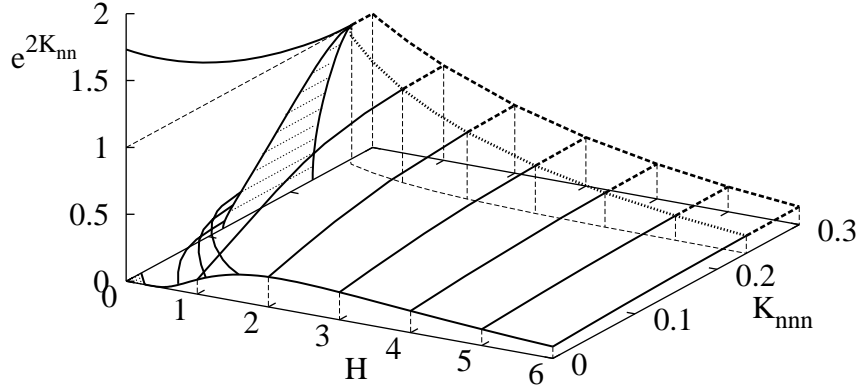}
\caption[xxx]
{The complete phase diagram in the three-parameter space 
($H, K_{\rm nnn},e^{2K_{\rm nn}}$). 
The solid lines denote second-order phase transitions, and the heavy 
dotted line is the tricritical line separating the three-state Potts
critical sheet from the first-order sheet which, shown by heavy dashed
lines. 
The three-state Potts critical surface is believed to connect to the
$e^{2K_{\rm nn}}=0$ plane at the KT line near the origin, and at the
$K_{\rm nnn}$ axis until the appearance of the critical phase.
The algebraic phases for $H=0$ and for $ T=0$ are lightly shaded,
and the thin dashed lines are projection lines added for clarity.
The error margins are at most of the same order as the thickness of
the lines.}  
\label{fig7}
\end{figure}

\begin{figure}
\includegraphics{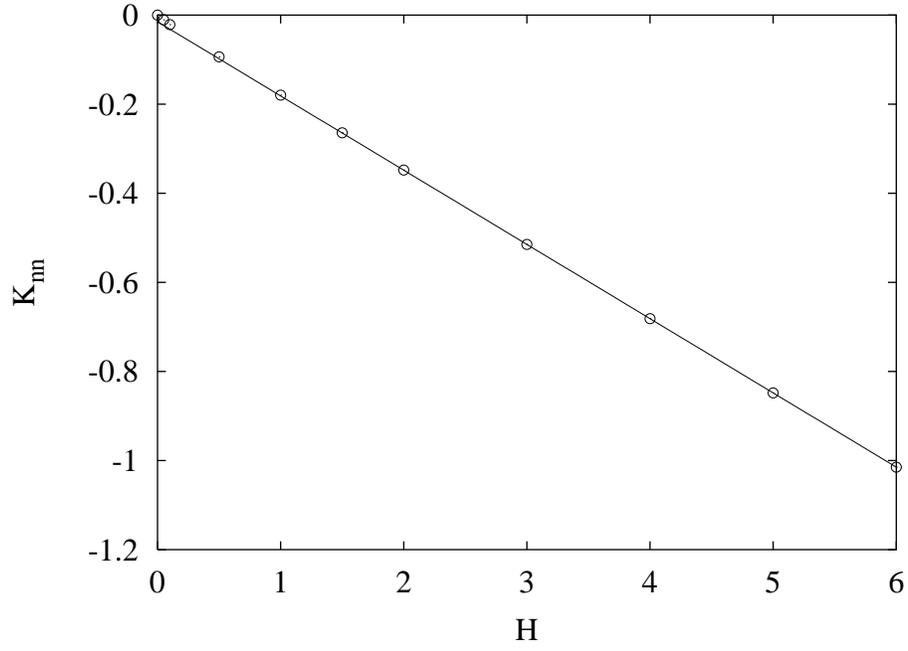}
\caption[xxx]
{The tricritical line shown as $K_{\rm nn}$ versus $H$.
The numerically determined tricritical points are shown as circles, and 
the solid line represents the
tricritical lattice-gas limit as $K_{\rm nn}=-H/6-0.01481$.}
\label{trifig}
\end{figure}

Renormalization analysis predicts that the uniform magnetic field
$H$ is relevant, except for a small range $2\leq g_R \leq 9/4$. 
Thus the plane $H=0$ qualifies as a possible locus of new universality
classes, in line with the existence of a critical phase such as 
predicted by the renormalization scenario and confirmed numerically.
We finally note that the renormalization equations for the KT
transitions imply that the line of KT transitions, as shown in
Fig.~\ref{fig2} on the left hand boundary of the critical phase, 
should come in as a straight line on the horizontal axis, in contrast
with the numerical results which there display a small part with a
sudden curvature. We believe that this is a finite-size effect,
explained by the same renormalization
equations, which involve the marginally irrelevant temperature field
parametrizing the line of KT transitions. This scaling field generates
slowly converging finite-size corrections. This field and its
associated finite-size effects vanish at $K_{\rm nn}=-\infty$.
\begin{acknowledgments}
We are indebted to Jouke R. Heringa for his contribution to the
development of the geometric cluster algorithm used in this work, and
to Bernard Nienhuis for valuable discussions.
\end{acknowledgments}

\begin{table}[ht]
\caption{\label{table1} Fitted results for the extrapolated average slope
$S_{\infty} \approx {\rm d}X_m /{\rm d}K_{\rm nnn}$ in the algebraic phase.
The last column shows the the exponent $y_c$ of finite-size correction.
The increase of $|S_{\infty}|$ with $K_{\rm nnn}$ corresponds with the
narrowing of the algebraic phase when the decoupling point $K_{\rm nn}=0$
is approached.
The intervals of $K_{\rm nnn}$ in which the average slopes are calculated
are listed in the second column.}
\vskip 2mm
\begin{tabular}{|c|c|c|c|}
\hline
\hline
 $K_{\rm nn}$ & $K_{\rm nnn}$ & $S_{\infty}$ & $y_c$ \\
\hline
-$\infty$ & 0.18 -  0.20 &$-0.59$ (3) &$-1.1$ (2) \\ 
$-0.6$    & 0.18 -  0.20 &$-0.78$ (2) &$-1.2$ (2) \\
$-0.4$    & 0.18 -  0.22 &$-1.20$ (8) &$-0.7$ (2) \\ 
$-0.2$    & 0.21 -  0.22 &$-3.3 $ (5) &$-0.3$ (1) \\
$-0.1$    & 0.23 -  0.25 &$-5.0 $ (10)&$-0.2$ (1) \\ 
\hline
\hline
\end{tabular}
\end{table}

\begin{table}
\caption{\label{table2} Tricritical points as obtained by the transfer 
matrix method  for several values of $H$. The decoupling point
$K_{\rm nn}=0$ is included here as the end point of the tricritical line,
although it does itself not belong to the tricritical three-state Potts
universality class. }
\vskip 2mm
\begin{tabular}{|c|c|c|}
\hline
\hline
 $H$    &  $K_{\rm nn}$ & $K_{\rm nnn}$   \\
\hline
  0.00  &$ 0.0000 $ (0) &  $\ln(3)/4$  (0) \\
  0.05  &$-0.0107 $ (12)&  0.269     (1) \\
  0.10  &$-0.0214 $ (10)&  0.2654    (5) \\
  0.5   &$-0.0937 $  (5)&  0.2572    (5) \\
  1.0   &$-0.1799 $  (2)&  0.2500    (2) \\
  1.5   &$-0.2644 $  (2)&  0.2452    (2) \\
  2.0   &$-0.3481 $  (2)&  0.2421    (2) \\
  3.0   &$-0.5150 $  (1)&  0.23845   (8) \\
  4.0   &$-0.6816 $  (1)&  0.23678   (8) \\
  5.0   &$-0.84823$  (5)&  0.23599   (8) \\
  6.0   &$-1.01487$  (5)&  0.23560   (8) \\
\hline
\hline
\end{tabular}
\end{table}

\noindent

\end{document}